\documentclass[12pt]{iopart}

\usepackage{iopams}  

\usepackage{bm}

\usepackage{setstack}

\usepackage{graphicx} 
\usepackage[biblabel]{cite}

\usepackage{xcolor}

\begin{document}

\title{Entanglement and Optimal Timing in Discriminating Quantum Dynamical Processes}

\author{Massimiliano F Sacchi$^{1,2}$}
\address{$^1$ CNR - Istituto di Fotonica e Nanotecnologie, Piazza Leonardo da Vinci 32, I-20133, Milano, Italy}
\address{$^2$ QUIT Group, Dipartimento di Fisica, Universit\`a degli Studi di Pavia, Via Agostino Bassi 6, I-27100 Pavia, Italy}


\begin{abstract}
I investigate the problem of optimally discriminating between two open
quantum dynamical processes in a single-shot scenario, with the goal
of minimizing the error probability of identification. This task
involves optimising both the input state---potentially entangled with
an ancillary system that remains isolated from the dynamics---and the
time at which the resulting time-dependent quantum channels, induced
by the two distinct dynamical maps, becomes most distinguishable. To
illustrate the complexity and richness of this problem, I focus on
Pauli dynamical maps and their associated families of time-dependent
Pauli channels. I identify a regime in which separable strategies
require waiting indefinitely for the dynamics to reach the stationary
state, whereas entangled input states enable optimal discrimination at
a finite time, with a strict reduction in error probability. These
results highlight the crucial interplay between entanglement and
timing in enhancing the distinguishability of quantum dynamical
processes.
\end{abstract}

%
\vspace{2pc}

\maketitle

\section{Introduction}
Discriminating among alternative hypotheses is a fundamental task in
both foundational and applied aspects of quantum information
theory. The most prominent example is the problem of minimizing the
error probability in quantum state discrimination, originally
formulated by Helstrom \cite{hel}. This challenge has since been
extensively explored in a variety of settings
\cite{unam,unam2,unam3,unam4,walg,virm,rev12,qifr,qifr2,qifr3},
including the discrimination of unitary transformations
\cite{CPR,aci,lop} and general quantum channels
\cite{qi1,qi2,chiri,naka,osk}. In particular, for minimum-error
discrimination between two quantum channels, it is well established
that the use of entangled input states can offer a strict
advantage---even in highly noisy scenarios, such as when dealing with
entanglement-breaking channels \cite{qi2}. This key insight has led to the
development of quantum illumination protocols
\cite{illu,illu2,illu3,illu4}, where entangled states of light enhance
the detection of weak signals embedded in noisy environments.

In this work, I focus on the problem of distinguishing between two
known open quantum dynamical processes, aiming to minimize the error
probability in a single-shot setting. I formulate the problem as
jointly optimising both the input state---which may be entangled with
an ancillary system that remains isolated from the dynamics---and the
time at which the time-dependent quantum channels, generated by the
two distinct processes, are most distinguishable. The analysis
presented here highlights the deep connection between entanglement,
temporal dynamics, and optimal discrimination, offering new insights
into the role of quantum resources in hypothesis testing.

To illustrate the complexity of the problem and the diversity of its
solutions, I present a detailed analysis of Pauli dynamical maps and
their corresponding families of time-dependent Pauli channels. These
channels model relevant noise processes in qubit systems and serve as
an instructive testbed for exploring optimal discrimination strategies
\cite{qiP,qif}. I show that in certain cases, optimal discrimination
without entanglement requires an indefinitely long waiting
time---until the process reaches its stationary state. In contrast,
the use of entangled input states enables discrimination at a finite
optimal time, along with a significant reduction in the error
probability. These findings underscore the crucial role of
entanglement and precise timing in enhancing the distinguishability of
quantum dynamical processes.

Finally, I note that prior studies using quantum probes to
discriminate between thermal baths, characterized by different
temperatures \cite{rud,mga} or statistical properties
\cite{cav1,cav2}, can be viewed as specific instances of the general
problem of distinguishing open quantum dynamics.

\section{Discriminating quantum dynamical processes}
A quantum dynamical process is governed by a generator ${\cal L}_t$
which dictates the evolution of the system density matrix $\rho (t)$
through the differential equation $\partial _t \rho (t)= {\cal L}_t
\rho (t)$. Under suitable physical conditions, the generator ${\cal
  L}_t$ can assume particularly tractable forms as, for example,
${\cal L}_t=- i[H_t, \cdot]$ for Hamiltonian dynamics, or a GKSL-like
form for Markovian divisible evolution \cite{pb,rh}. Anyway, to
represent a physically valid evolution, at any time $t$ the associated
map ${\cal E}_t ={\cal T} \exp (\int _{0}^t {\cal L}_s ds)$, defined
via the time-ordered exponential of ${\cal L}_t$, must be a quantum
channel---i.e. a trace-preserving completely positive map
\cite{pb,rh}. In fact, this map ${\cal E}_t$ provides the quantum
state $\rho (t)$ at time $t$ from the initial state $\rho(0)$ as $\rho
(t)={\cal E}_t (\rho (0))$.

Now consider a scenario in which the evolution of a quantum system is
known to be governed by one of two candidate dynamical processes,
generated by ${\cal L}_t^{(1)}$ and ${\cal L}_t^{(2)}$, occurring with
prior probabilities $q_1$ and $q_2=1-q_1$, respectively. The problem
of distinguishing between these two dynamics reduces to identifying
the optimal time $t^*$ at which a discrimination test between the
corresponding time-dependent quantum channels ${\cal E}^{(1)}_t$ and
${\cal E}_t^{(2)}$ minimises the probability of error. This
optimisation may include the use of entangled input states to enhance
distinguishability.

To frame this problem, I briefly recall the main results from
Ref. \cite{qi1} on minimum-error discrimination between two generic
quantum channels ${\cal E}^{(1)}$ and ${\cal E}^{(2)}$, given with
prior probabilities $q_1$ and $q_2$. This problem is formulated by
seeking the optimal input state $\rho $ on the Hilbert space $\cal H$
such that the error probability in discriminating the corresponding
output states ${\cal E}^{(1)} (\rho )$ and ${\cal E}^{(2)}(\rho )$ is
minimised.  If side entanglement is allowed, the output states to be
distinguished take the form $({\cal E}^{(1)}\otimes {\cal I} ) \rho $
and $({\cal E}^{(2)}\otimes {\cal I}) \rho $, where the input $\rho $
is generally a bipartite state on ${\cal H}\otimes {\cal K}$, and the
quantum channels act solely just on the first subsystem, while the
trivial identity map ${\cal I}$ acts on the second. Notably, the use
of entanglement can strictly enhance discrimination, even in highly
noisy scenarios \cite{qi2}.

The foundational result by Helstrom \cite{hel} states that the minimum
error probability $p'_E$ in optimally discriminating between two
quantum states $\rho _1$ and $\rho _2$, given with prior probabilities
$q_1$ and $q_2$, is given by
\begin{eqnarray}
p'_E= \left (1 -\Vert q_1 \rho_ 1 -q _2 \rho _2 \Vert _1 \right
)/2\;,\label{pest}
\end{eqnarray}
where $\Vert A \Vert _1= \hbox{Tr}\sqrt{A^\dag A} $ denotes the trace
norm of $A$.  This result inherently accounts for the corresponding optimal (binary and orthogonal)
measurement.

\par Then, for quantum channel discrimination with no use of
entanglement, the minimum error probability $\tilde p_E$ is given by
\begin{eqnarray}
\tilde p_E 
=
(1- \max _{\rho} 
\Vert q_1 {\cal E}^{(1)}
(\rho )- q_2{\cal E}^{(2)} (\rho )\Vert  _1 )/2
\,,\label{peno}
\end{eqnarray}
where $\rho $ ranges over all density matrices on $\cal H$.  On the
other hand, when entangled input states are allowed, the minimum error
probability $p_E$ is given by
\begin{eqnarray}
p_E =(1- \max _{\rho }
\Vert q_1 ({\cal E}^{(1)} \otimes {\cal I})
\rho - q_2 ({\cal E}^{(2)}\otimes {\cal I})\rho \Vert _1  )/2
\,,\label{pesi}
\end{eqnarray}
where now $\rho $ is a density matrix on  ${\cal H}\otimes {\cal K}$.  The
maximum of the trace norm in Eq. (\ref{pesi}) is known as the norm of
complete boundedness (or diamond norm). In fact, for
finite-dimensional Hilbert space, one can simply take ${\cal K}={\cal
  H}$ \cite{paulsen,diam}.  Moreover, due to the convexity of the trace
norm, in both Eqs. (\ref{peno}) and (\ref{pesi}) the maximum can be
searched for among pure states.

It follows that the error probabilities in discriminating between the
two dynamical maps ${\cal L}_t^{(1)}$ and ${\cal L}_t^{(2)}$ at time
$t$ can be obtained from Eqs. (\ref{peno}) and (\ref{pesi}) by
replacing ${\cal E}^{(i)} $ with ${\cal E}_t^{(i)} ={\cal T} \exp
(\int _{0}^t {\cal L} ^{(i)}_s ds)$. In this way, I promote $p_E$ (and
$\tilde p_E$) to time dependent functions, whose infimum over $t>0$
provides the ultimate minimum error probability $p_E^*$ (and $\tilde
p_E^*$) for discriminating between the dynamical maps, with (and
without) use of side entanglement. I remark that when both dynamical
processes are Hamiltonian, i.e. ${\cal L}^{(i)}_t=- i[H^{(i)}_t,
  \cdot]$, entanglement never improves discrimination, since the two
corresponding channels ${\cal E}_t^{(i)}$ are unitary \cite{CPR,aci,lop}.

\section{The case of two Pauli dynamical maps}
In the following  I consider the explicit case of two Pauli dynamical maps for
qubits, namely 
\begin{eqnarray}
  \partial _t \rho (t)= {\cal L} ^{(i)}\rho (t)
  = \sum _{k=1}^3 \gamma
^{(i)} _k
[\sigma _k \rho (t) \sigma _k - \rho (t)]
\;.\label{meq}
\end{eqnarray}
where $\{\sigma _1\,,\sigma _2\,,\sigma _3 \}= \{ \sigma _x\,,\sigma
_y\,,\sigma _z\}$ denote the Pauli matrices, and $\bm{\gamma
}^{(i)}\equiv \{\gamma ^{(i)}_x\,, \gamma ^{(i)}_y\,, \gamma
^{(i)}_z\}$ represent the vector of the pertaining decay rates.  These
maps are purely dissipative, have time-independent generators, and
describe two distinct semi-group Markovian dynamics. The
solutions ${\cal E}_t ^{(i)}=e^{{\cal L}^{(i)}t}$ of Eq. (\ref{meq})
define two sets of Pauli channels ${\cal
  E}_t ^{(1)}$ and ${\cal E}_t^{(2)}$, namely
\begin{eqnarray}
  \rho ^{(i)}(t) = {\cal E}_t ^{(i)} (\rho (0))= \sum_{k=0}^3 p_k ^{(i)}(t)
  \sigma _k \rho (0) \sigma _k\;,\label{pc}
\end{eqnarray}
where I introduced $\sigma _0 \equiv  I$ as the $2\times 2$ identity
matrix along with the two time-dependent probability vectors $\{p^{(i)}_k(t)
\}$. These probabilities are related to the decay rates $\{\gamma
^{(i)}_k\}$ in Eq. (\ref{meq}) by the relations \cite{CW}
\begin{eqnarray}
p_k ^{(i)}(t)= \frac 14 \sum _{l=0}^3  H_{kl} A_l ^{(i)} (t)\;,\label{cw}
\end{eqnarray}
where $A_l ^{(i)}(t)$ are the components of the vectors
\begin{equation}
  A^{(i)}(t)
  =
  \left(
  \begin{array}{c}
    1 \\
    e^{-2 (\gamma _2 ^{(i)}+ \gamma _3^{(i)})t}\\
    e^{-2 (\gamma _1 ^{(i)}+ \gamma _3 ^{(i)})t}\\
    e^{-2 (\gamma _1 ^{(i)}  + \gamma _2 ^{(i)})t}
    \end{array}
    \right)
\;,
\end{equation}
and $H_{kl}$ denote  the elements of the Hadamard matrix
\begin{equation}
  H=
  \left( \begin{array}{rrrr}
    1 & 1 & 1 & 1 \\
    1 & 1 & -1 & -1 \\
    1 & -1 & 1 &- 1 \\
    1 & -1 & -1 & 1 
  \end{array} \right) \;.
\end{equation}  
\par I now recall from Refs. \cite{qi1,qiP} the expressions for the error probabilities (\ref{peno}) and
(\ref{pesi}) in the case of two Pauli channels ${\cal E}^{(i)} (\rho) = \sum_{k=0}^3 p_k ^{(i)}
\sigma _k \rho \sigma _k$. By defining
\begin{eqnarray}
r_k \equiv  q_1 p_k ^{(1)} - q_2 p_k^{(2)}\;,\label{rkk}
\end{eqnarray}
one has 
\begin{eqnarray}
  \tilde p_E = (1 -M)/2\;,
\end{eqnarray}
where
\begin{eqnarray}
M=  \max \{ &&|r_0  + r_3|+|r_1 +r_2 |\,, 
|r_0 +r_1 |+|r_2+r_3|\,,  \nonumber \\& & 
|r_0 + r_2 |+|r_1 +r_3 |  \}
\;.\label{emme} 
\end{eqnarray}
The three cases compared inside the curly brackets corresponds to feeding the
unknown channel with an eigenstate of $\sigma _z$, $\sigma _x$, and $\sigma _y$,
respectively. 
\par On the other hand, the error probability obtained when using side 
entanglement is given by \cite{qi1,qiP}
\begin{eqnarray}
p_E = (1 -\textstyle \sum_{k=0}^3 |r_k |)/2\;,\label{peu}
\end{eqnarray}
and is achieved by using an arbitrary two-qubit maximally
entangled input state. Entanglement strictly improves the discrimination,
i.e. $p_E < \tilde p_E$, iff $\Pi _{k=0}^3 r_k <0$.

With these results, I have provided all the necessary ingredients to
solve the problem of the optimal discrimination between the two
dynamical maps ${\cal L}^{(1)}$ and ${\cal L}^{(2)}$ in
Eq. (\ref{meq}), which generate the two possible evolutions of
Eq. (\ref{pc}). The error probabilities for discrimination (with and
without side entanglement) at time $t$ can now be evaluated by
promoting the values $r_k $ in Eq. (\ref{rkk}) to time-dependent
functions, expressed in terms of the time-dependent probabilities $p_k
^{(1)}(t)$ and $p_k ^{(2)}(t)$ obtained by Eq. (\ref{cw}).  The resulting
time-dependent minimum error probabilities, that henceforth I denote
as $p_E(t)$ and $\tilde p_E(t)$, can then be further optimised over
time $t$, yielding the ultimate minimum error probabilities $p_E^*$
and $\tilde p_E^*$ for distinguishing between the dynamical maps
${\cal L}^{(1)}$ and ${\cal L}^{(2)}$.

Hereafter, I present explicit solutions for several 
representative cases. For simplicity, I assume equal prior
probabilities, i.e., $q_1 = q_2 =\frac 12$.
\subsection{Two dephasing processes along the same direction}  
Consider two dephashing processes aligned along the same direction,
e.g., corresponding to the eigenbasis of $\sigma _z$. This implies that in Eq. (\ref{meq})
we take $\bm{\gamma }^{(i)}=\{0,0,\gamma ^{(i)}\}$. The 
dynamics described by Eq. (\ref{pc}) then corresponds to two
dephasing channels with
\begin{eqnarray}
  &&p_0^{(i)}(t)= (1 +e ^{-2 \gamma ^{(i)}t})/2  \\
&&p_1^{(i)}(t)= p_2^{(i)}(t)= 0  \\
    &&p_3^{(i)}(t)= (1 -e ^{-2 \gamma ^{(i)}t})/2
  \;.
\end{eqnarray}
Consequently, we obtain  
\begin{eqnarray}
  p_E (t)=\tilde p_E(t) = \frac 12 -\frac 14 \left | e^{-2 \gamma ^{(1)}t}-e^{-2 \gamma ^{(2)}t}
  \right |\;.\label{petdeph}
\end{eqnarray}
Thus, entanglement provides no advantage in discrimination at any time. By solving $\partial _t p_E (t)=0$ one finds
the optimal time  $t^*$ for comparing the two
processes, namely 
$t^*= \frac{1}{2(\gamma ^{(1)}- \gamma  ^{(2)})} \ln  
\frac{\gamma ^{(1)}}{\gamma
  ^{(2)}}$, with the corresponding minimum error probability
\begin{eqnarray}
  p_E^*=\tilde p_E^*=
  \frac 12 -\frac 14 \left(  \frac{\gamma ^{(1)}}{\gamma
  ^{(2)}}\right )^{\frac {\gamma ^{(1)}}
  {\gamma ^{(2)}- \gamma  ^{(1)}}}
\left | \frac {\gamma ^{(1)}}{\gamma ^{(2)}}- 1 \right |
\;.\label{pdephug}
\end{eqnarray}
This result shows that $p_E^*$ depends only on the ratio
of the decay rates. Figure 1 illustrates the error probability $p_E (t)$
for distinguishing a dephasing process with ${\gamma }^{(1)} =1$
from those with $\gamma ^{(2)} =0.25$,  $0.5$, and $4.0$. 
\begin{figure}[ht]
\begin{center}
  \includegraphics[scale=0.6]{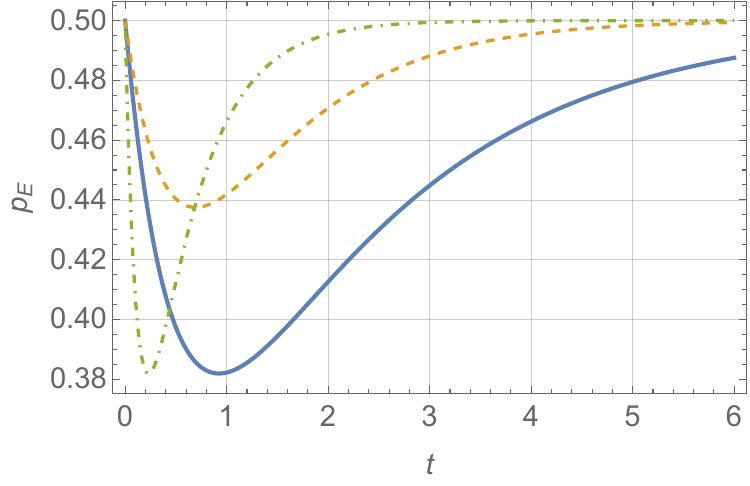}
\end{center}  
  \caption{Minimum error probability for discriminating at time $t$
    a dephasing process ${\bm\gamma }^{(1)} =(0,0,1)$ from those
    with ${\bm\gamma }^{(2)} =(0,0,\gamma ^{(2)})$, where $\gamma
    ^{(2)}=0.25$, $0.5$, and $4.0$ (solid, dashed, and dot-dashed
    lines, respectively). Entanglement provides no
    advantage at any time.}\end{figure}
\subsection{Two dephasing processes along orthogonal directions} 
Now, consider two dephasing processes aligned along orthogonal
directions, e.g., corresponding to the mutually unbiased bases of
$\sigma _z$ and $\sigma _x$. This means solving the case with 
$\bm{\gamma }^{(1)}=\{0,0,\gamma ^{(1)}\}$ and $\bm{\gamma
}^{(2)}=\{\gamma ^{(2)},0,0\}$. As before, the dynamics
corresponds to dephasing channels, but now the second channel dephases
along $\sigma _x$, namely
\begin{eqnarray}
  &&p_0^{(2)}(t)= (1 +e ^{-2 \gamma ^{(2)}t})/2  \\
    &&p_1^{(2)}(t)= (1 -e ^{-2 \gamma ^{(2)}t})/2  \\
  &&p_2^{(2)}(t)= p_3^{(2)}(t)= 0  
  \;.
\end{eqnarray}
Thus, we obtain 
\begin{eqnarray}
  p_E (t)=\tilde p_E(t) = \frac 14 \left ( 1+ e^{-2 \max
    \{ \gamma^{(1)},\gamma^{(2)} \}t }\right ) \;.\label{petdeph2}
\end{eqnarray}
Again, entanglement does not enhance discrimination at any time. 
Clearly, the optimal discrimination occurs at $t\rightarrow +\infty$, where $p_E^*=\tilde p_E^*=
\frac 14$.
\subsection{Two coplanar decaying processes} 
Next, consider two processes with $\bm{\gamma
}^{(i)}=\{\gamma ^{(i)},\gamma ^{(i)},0\}$.  Each
process provides a set of Pauli channels with the time-dependent probabilities
\begin{eqnarray}
  &&p_0^{(i)}(t)= e ^{-2 \gamma ^{(i)}t} \cosh ^2 (\gamma ^{(i)}t) \\
  &&p_1^{(i)}(t)=p_2^{(i)}(t)=(1- e^{-4 \gamma ^{(i)}t})/4  \\
  &&   p_3^{(i)}(t)=  e ^{-2 \gamma ^{(i)}t} \sinh ^2 (\gamma ^{(i)}t) 
\;.
\end{eqnarray}
In this case, entanglement improves discrimination at any time. Specifically, we have
\begin{eqnarray}
\tilde p_E(t)=\frac 12 -   
  \frac 14 \max \{ 
|e^{-2 \gamma
  ^{(1)}t}-e^{-2 \gamma ^{(2)}t} |,
|e^{-4 \gamma
  ^{(1)}t}-e^{-4 \gamma ^{(2)}t} | \}
\label{cop}\;,
\end{eqnarray}
while
\begin{eqnarray}
p_E (t)=\frac 12 -   \frac 14 | e^{-2 \gamma
    ^{(1)}t}-e^{-2 \gamma ^{(2)}t} | 
-  \frac 18 |e^{-4 \gamma
    ^{(1)}t}-e^{-4 \gamma ^{(2)}t} |
\;,\label{cop2}
\end{eqnarray}
and, clearly, $p_E(t) < \tilde p_E(t)$.  In the absence of side
entanglement, since $\tilde p_E (t)$ has two absolute minima, there
are two optimal times given by $t^* =\frac {\kappa }{\gamma
  ^{(1)}-\gamma ^{(2)}} \ln \frac {\gamma ^{(1)}}{\gamma ^{(2)}} $,
with $\kappa = 1/2$ or $1/4$. At both times, we have
\begin{eqnarray}
  \tilde p_E^*
=\frac 12 -\frac 14 \left(  \frac{\gamma ^{(1)}}{\gamma
  ^{(2)}}\right )^{\frac {\gamma ^{(1)}}
  {\gamma ^{(2)}- \gamma  ^{(1)}}}
\left | \frac {\gamma ^{(1)}}{\gamma ^{(2)}}- 1 \right |
  \,.
\end{eqnarray}
On the other hand, when using entanglement, there is a unique optimal
time $t^*$ such that $\partial _t p_E (t)=0$, which corresponds to the solution of the transcendental
equation
\begin{eqnarray}
  \gamma ^{(1)}(e^{-4 \gamma ^{(1)}t}+e^{-2 \gamma ^{(1)}t})=
  \gamma ^{(2)}(e^{-4 \gamma ^{(2)}t}+e^{-2 \gamma ^{(2)}t})
\;.
\end{eqnarray}
Figure 2 shows the error probabilities from Eqs. (\ref{cop}) and
(\ref{cop2}) for $\gamma ^{(1)}=1$ and $\gamma ^{(2)}=0.2$. The
ultimate minimum error probability, $p_E^*\simeq 0.308$, is achieved using
entanglement, with discrimination performed at the optimal time
$t^*\simeq 0.782$.
\begin{figure}[htb]
\begin{center}
  \includegraphics[scale=0.6]{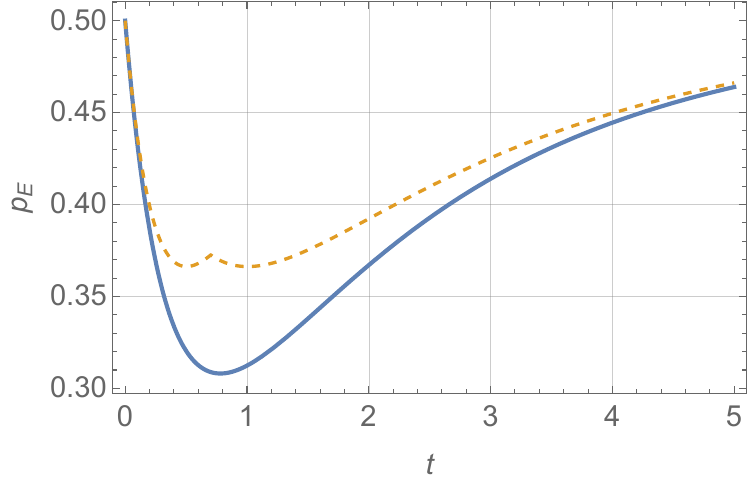}
  \end{center}
  \caption{Minimum error probabilities for discriminating between two
    coplanar decaying processes with ${\bm\gamma }^{(1)} =(1,1,0)$ and
    ${\bm\gamma }^{(2)} =(.2,.2,0)$ at time $t$, with and without
    entanglement (solid and dashed lines, respectively). Side
    entanglement strictly improves discrimination at any time.}
\end{figure}
\subsection{Two depolarising processes}
In a depolarising process, the
decay rates have equal and constant components. Thus, we consider the
case with $\bm{\gamma }^{(i)}=\{\gamma ^{(i)},\gamma ^{(i)},\gamma
^{(i)}\}$. The dynamics of each process gives a set depolarising
channels with the time-dependent probabilities
\begin{eqnarray}
  &&p_0^{(i)}(t)= (1+3 e^{-4 \gamma ^{(i)}t})/4 \\
  &&p_1^{(i)}(t)=p_2^{(i)}(t)=  p_3^{(i)}(t)= (1-  e^{-4 \gamma ^{(i)}t}) /4
\;.
\end{eqnarray}
Thus, we obtain 
\begin{eqnarray}
&&\tilde p_E (t)=\frac 12 -   \frac 14 | e^{-4 \gamma
    ^{(1)}t}-e^{-4 \gamma ^{(2)}t} | \;,\\& &
  p_E (t)=\frac 12 -   \frac 38 | e^{-4 \gamma
    ^{(1)}t}-e^{-4 \gamma ^{(2)}t} | \;.
\end{eqnarray}
Clearly, entanglement enhances discrimination at any time. Figure 3
illustrates this for the specific values $\gamma ^{(1)} =1$ and
$\gamma ^{(2)}=0.2$.
\begin{figure}[ht]
  \begin{center}
    \includegraphics[scale=0.6]{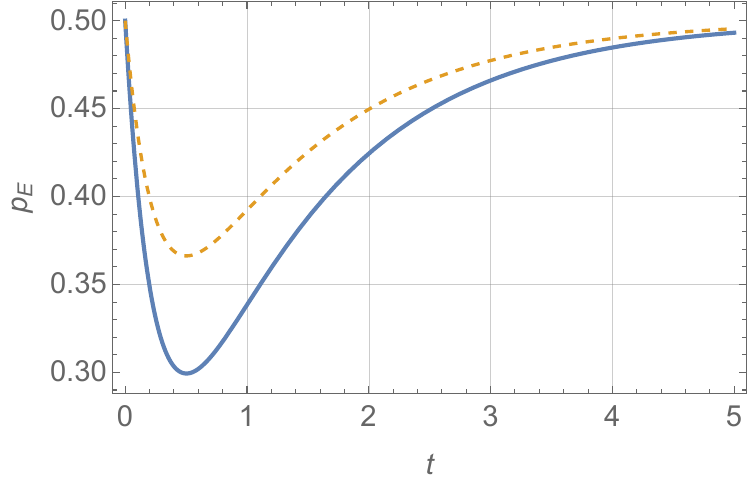}
    \end{center}
  \caption{Minimum error probability for discriminating 
    two depolarising processes at time $t$, with ${\bm\gamma }^{(1)} =(1,1,1)$ and
    ${\bm\gamma }^{(2)}=(0.2,0.2,0.2)$, with and without entanglement
    assistance (solid and dashed lines, respectively).}
\end{figure}
\par\noindent Notice that both $\tilde p _E(t)$ and $p_E(t)$ reach their minimum
at the same optimal time 
\begin{eqnarray}
t^*= \frac{1}{4 (\gamma ^{(1)}- \gamma  ^{(2)})} \ln  
\frac{\gamma ^{(1)}}{\gamma
  ^{(2)}}
  \,,\label{tss}
\end{eqnarray}
with the corresponding ultimate minimum error probabilities given by 
\begin{eqnarray}
  && \tilde p_E^*= 
\frac 12 -\frac 14 \left(  \frac{\gamma ^{(1)}}{\gamma
  ^{(2)}}\right )^{\frac {\gamma ^{(1)}}
  {\gamma ^{(2)}- \gamma  ^{(1)}}}
\left | \frac {\gamma ^{(1)}}{\gamma ^{(2)}}- 1 \right | \,,\label{ultii}
\\&&
p_E^*= 
\frac 12 -\frac 38 \left(  \frac{\gamma ^{(1)}}{\gamma
  ^{(2)}}\right )^{\frac {\gamma ^{(1)}}
  {\gamma ^{(2)}- \gamma  ^{(1)}}}
\left | \frac {\gamma ^{(1)}}{\gamma ^{(2)}}- 1 \right |   
\,.\label{ulti}
\end{eqnarray}
These error probabilities are shown in Fig. 4 as a function of the ratio of the decay parameters
$\gamma ^{(1)}$ and $\gamma ^{(2)}$.
\begin{figure}[ht]
\begin{center}
  \includegraphics[scale=0.6]{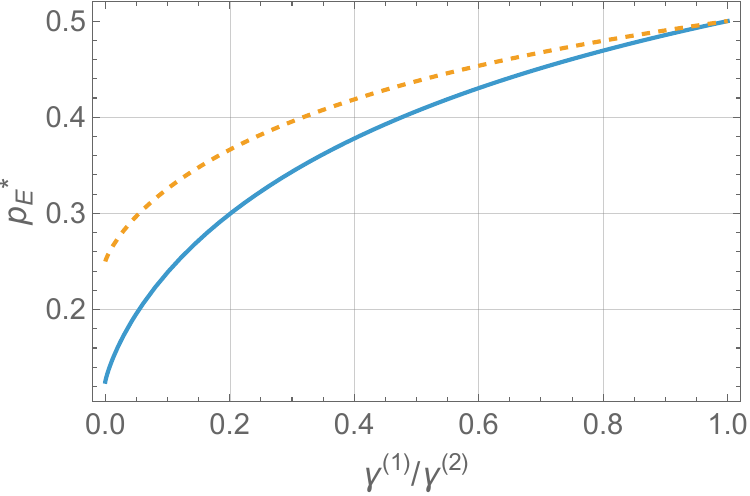} 
\end{center}
  \caption{Ultimate minimum error probabilities in discriminating
    between two depolarising processes as a function of the ratio 
    of the decay rates, with and
    without entanglement assistance (solid and dashed lines,
    respectively). In both cases, the optimal discrimination time is
    given by Eq. (\ref{tss}).}
\end{figure}
\subsection{A depolarising and a dephasing process} 
Finally, let us consider the problem of distinguishing between a depolarising
process, $\bm{\gamma }^{(1)}=\{\gamma ^{(1)},\gamma ^{(1)},\gamma
^{(1)}\}$, and a  dephasing process, $\bm{\gamma }^{(2)}=\{0,0,\gamma
^{(2)}\}$. For a strategy with no use of entanglement the
error probability is given by
\begin{eqnarray}
\tilde p_E(t)=\frac 12 -   
  \frac 14 \max \{  1 -e^{-4 \gamma ^{(1)}t} ,
|e^{-4 \gamma
  ^{(1)}t}-e^{-2 \gamma ^{(2)}t} | \} \label{pettt} \;.
\end{eqnarray}
In this case, the infimum is achieved in the limit 
$t\rightarrow +\infty$, where $\tilde p_E^*=\frac 14$.

\par\noindent By utilizing entanglement, the error probability is instead given by
\begin{eqnarray}
p_E (t)&&=\frac 12 -   \frac 18 (1 -e^{-4 \gamma
    ^{(1)}t})
-  \frac {1}{16} |1 -3 e^{-4 \gamma
  ^{(1)}t} + 2 e^{-2 \gamma ^{(2)}t} |
\nonumber \\& &
-  \frac {1}{16} |1 + e^{-4 \gamma
  ^{(1)}t}  -  2 e^{-2 \gamma ^{(2)}t}|
\;.\label{pett}
\end{eqnarray}
The comparison between these two strategies is more intricate than in
previous cases. By analyzing the two
functions (\ref{pettt}) and (\ref{pett}), we find that they are equal $p_E(t)=\tilde
p_E(t)=\frac 14 (1+e^{-4 \gamma ^{(1)}}t)$ at all times when 
\begin{eqnarray}
3 e^{-4 \gamma ^{(1)}t} -1 \leq 2 e^{-2 \gamma ^{(2)}t} \leq e^{-4 \gamma
  ^{(1)}t} +1
\;.\label{cont}
\end{eqnarray}
To determine whether entanglement provides a strict advantage, we must
identify a time $t^*$ at which $p_E(t)$ reaches a minimum value lower than $\frac
14$. Notice that the condition $p_E(t)\leq \frac 14$ can be satisfied in the region
\begin{eqnarray}
e^{-2 \gamma ^{(2)}t} \geq (3 e^{-4 \gamma   ^{(1)}t}
  +1)/2
\;,
\end{eqnarray}
where
\begin{eqnarray}
p_E(t) = (3+ 3 e^{-4 \gamma   ^{(1)}t} -2 e^{-2 \gamma   ^{(2)}t})/8
\;.\label{pem}
\end{eqnarray}
In fact, numerical inspection shows that this condition holds as long as
$\frac{\gamma ^{(2)}}{\gamma ^{(1)}}\lesssim 0.3785$.  In this case
the minimum of $p_E(t)$ in Eq. (\ref{pem}) is reached at $t^*=
\frac{1}{4 \gamma ^{(1)} -2 \gamma ^{(2)}}\ln \frac {3\gamma
  ^{(1)}}{\gamma ^{(2)}}$, with a corresponding minimum error
probability
\begin{eqnarray}
p^*_E \!= \!\frac  38 \left [1 - \left ( \frac   {3\gamma
    ^{(1)}}{\gamma   ^{(2)}}  
  \right ) ^\frac {2\gamma   ^{(1)}}{\gamma   ^{(2)} -  2\gamma   ^{(1)}}
\left (\frac{2\gamma   ^{(1)}}{\gamma   ^{(2)}} -1\right )
      \right ]< \frac 14\,.
\end{eqnarray}
Thus, when $\frac{\gamma ^{(2)}}{\gamma ^{(1)}}\lesssim
0.3785$, entanglement provides a strict advantage in discrimination at a
finite time. Figure 5 illustrates this threshold, where the benefits
of entanglement are clearly visible.
\begin{figure}[ht]
\begin{center}
  \includegraphics[scale=0.5]{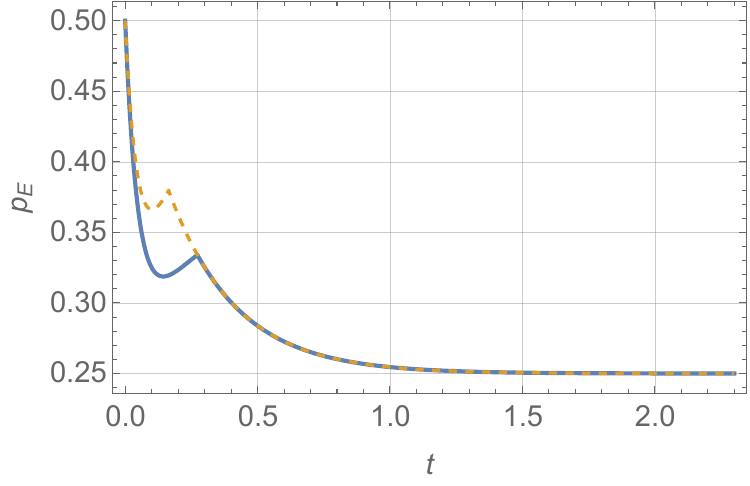} 
  \includegraphics[scale=0.5]{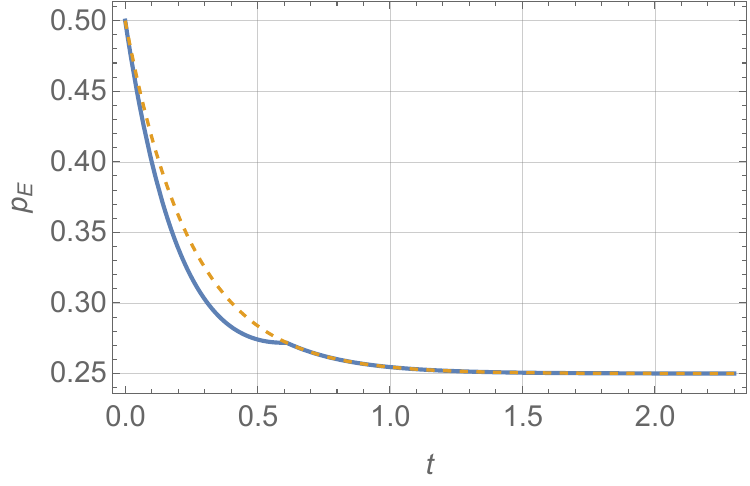}
  \includegraphics[scale=0.5]{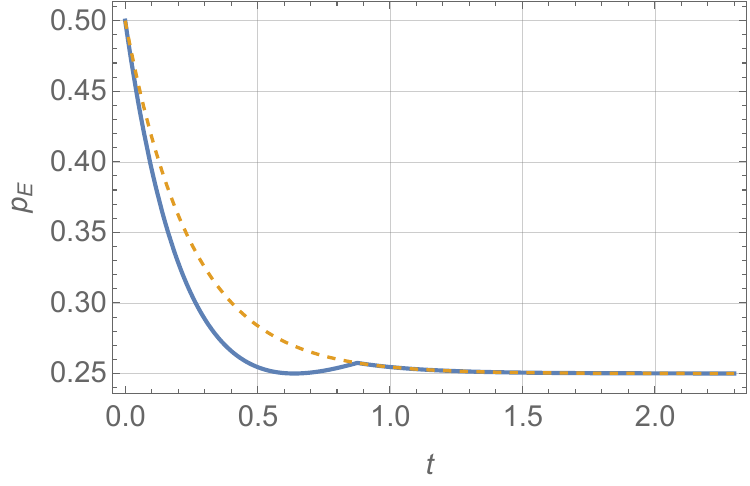}
\includegraphics[scale=0.5]{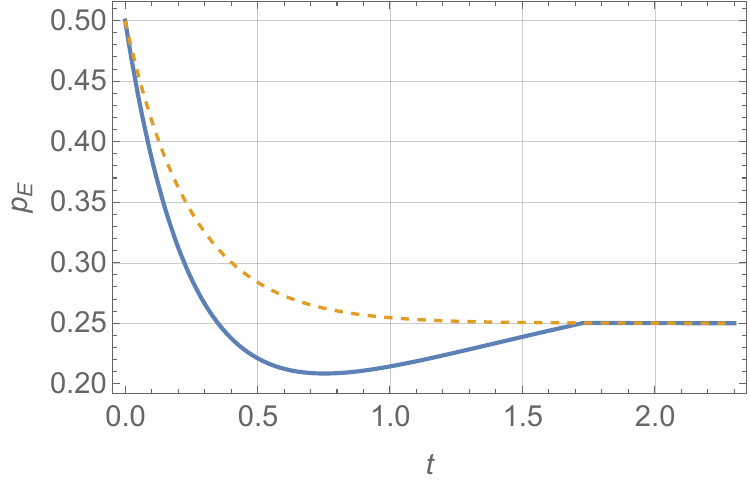}
\end{center}
\caption{Minimum error probability for discriminating the
    depolarising process ${\bm\gamma }^{(1)} =(1,1,1)$ from a dephasing
    process ${\bm\gamma }^{(2)} =(0,0,\gamma ^{(2)})$ with and
    without side entanglement
    (solid and dashed lines, respectively), for different values of
    $\gamma ^{(2)}$: $10$ (top left),  $0.5$ (top right), $0.3785$ (bottom left), and $0.2$
    (bottom right).}
\end{figure}
\section{Conclusions}
This work deepens our understanding of the discrimination of quantum
dynamical processes by uncovering the crucial interplay between
entanglement, timing, and optimal measurement strategies. By showing
that entangled input states can enable optimal discrimination at
finite times---whereas separable strategies may require waiting
indefinitely until stationarity---this study highlights the
operational value of quantum correlations in time-sensitive
tasks. These results establish a concrete foundation
for designing more efficient discrimination protocols and underscore
the broader utility of entanglement in dynamically evolving quantum
systems.

The findings naturally invite several directions for future research,
including extensions to more complex dynamical scenarios such as
processes governed by inequivalent Lindbladian generators or
exhibiting non-Markovian features. From a practical perspective, the
results have direct relevance to quantum metrology, error correction,
and sensing, where rapid and reliable identification of dynamical
behavior is often critical. In particular, in applications like
quantum illumination, optimal timing may correspond to optimal spatial
configurations—such as the distance to a target—further illustrating
the practical implications of temporal optimisation in real-world
quantum technologies.

 
\ack{This work has been sponsored by PRIN MUR Project 2022SW3RPY.}

\end{document}